\documentclass[ reprint, amsmath,amssymb,aps]{revtex4-2}
\usepackage{color}
\usepackage{natbib}
\usepackage{indentfirst}
\usepackage{graphicx}
\usepackage{dcolumn}
\usepackage{bm}
\usepackage[utf8]{inputenc}
\usepackage[T1]{fontenc}
\usepackage{mathptmx}
\usepackage{makecell}
\usepackage{multirow}

\begin{document}
\title{Why exercise builds muscles: Titin mechanosensing controls skeletal muscle growth under load}

\author{Neil Ibata and Eugene M. Terentjev}
\altaffiliation{Cavendish Laboratory, University of Cambridge, J.J. Thomson Avenue, Cambridge, CB3 0HE, U.K.}
\email{emt1000@cam.ac.uk}

\date{\today}

\begin{abstract}
Muscles sense internally generated and externally applied forces, responding to these in a coordinated hierarchical manner at different time scales. The center of the basic unit of the muscle, the sarcomeric M-band, is perfectly placed to sense the different types of load to which the muscle is subjected. In particular, the kinase domain (TK) of titin located at the M-band is a known candidate for mechanical signaling. Here, we develop the quantitative mathematical model that describes the kinetics of TK-based mechanosensitive signaling, and predicts trophic changes in response to exercise and rehabilitation regimes.
First, we build the kinetic model for TK conformational changes under force: opening, phosphorylation, signaling and autoinhibition. We find that TK opens as a metastable mechanosensitive switch, which naturally produces a much greater signal after high-load resistance exercise than an equally energetically costly endurance effort.
Next, in order for the model to be stable, give coherent predictions, in particular the lag following the onset of an exercise regime, we have to account for the associated kinetics of phosphate (carried by ATP), and for the non-linear dependence of protein synthesis rates on muscle fibre size. We suggest that the latter effect may occur via the steric inhibition of ribosome diffusion through the sieve-like myofilament lattice.
The full model yields a steady-state solution (homeostasis) for muscle cross-sectional area and tension,  and a quantitatively plausible hypertrophic response to training as well as atrophy following an extended reduction in tension. 
\end{abstract}


\maketitle

\section{Introduction}
Why does exercise build skeletal muscles, whereas long periods of immobility lead to muscle atrophy? The anecdotal evidence is clear, and the sports and rehabilitation medicine community has amassed a large amount of empirical knowledge on this topic. But the community has not as yet addressed and understood two key phenomena which underly hypertrophy and atrophy: how does the muscle `know' that it is being exercised (when it is certainly not the tactile sense, processed via the nervous system, that is at play in this), and how does it signal to provoke a morphological response to an increase or a lack of applied load? 
  In some communities, there is a perception that muscle grows after exercise due to its internal repair of micro-damage inflicted by the load. However, it is obvious that such an idea cannot be true, for several reasons: most of the `tissue repair' occurs by growing connective tissue, while we need an increase of intricately hierarchical myofilament structure, also this concept will not account for atrophy developing in microgravity or after extended bedrest. 

Here we develop a quantitative theoretical model which seeks to explain both of these processes. In order to be useful, the model must build on the relevant knowledge accumulated from studies of the anatomy and physiology of muscles, as well as the biological physics of molecular interactions and forces.

\begin{figure}
\centering
\includegraphics[width=0.99\columnwidth]{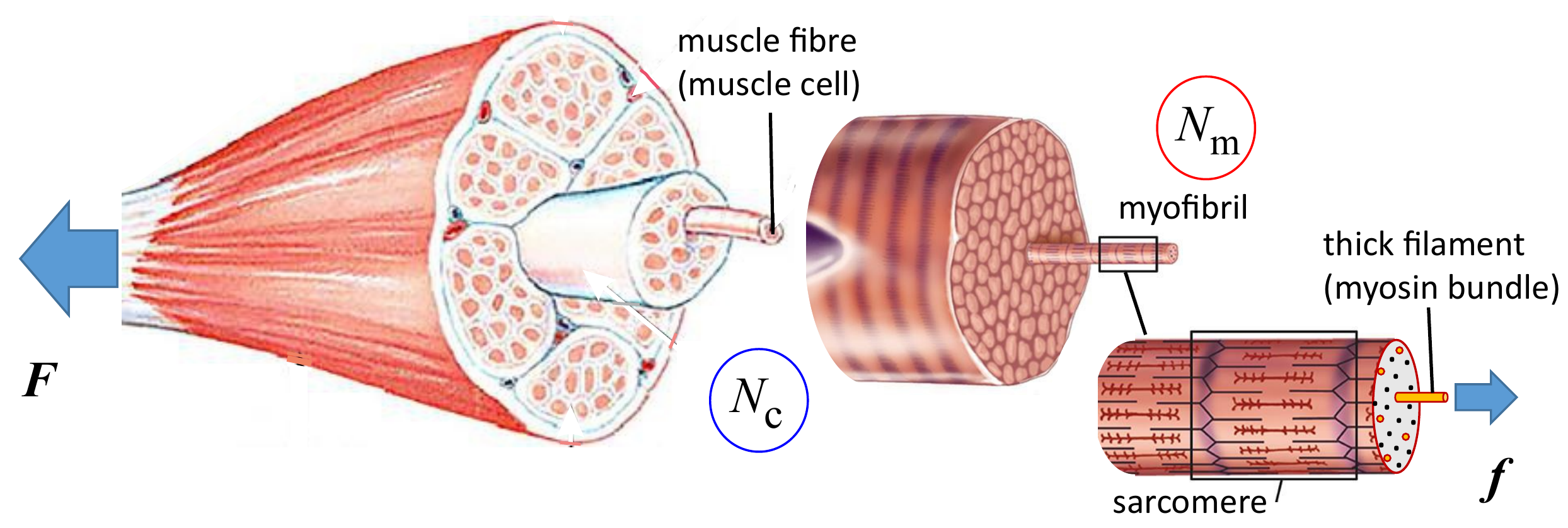}
\caption{The `textbook' hierarchy in the anatomy of skeletal muscle. The overall muscle is characterised by its cross-section area (CSA), which contains a certain number ($N_\mathrm{c}$) of mucle fibers (the muscle cells with multiple nuclei, or multinucleate myocytes). A given muscle has a nearly fixed number of myocytes: between $N_\mathrm{c}\approx 1000$ for the tensor tympani,
and $N_\mathrm{c} >$ 1,000,000 for large muscles: gastrocnemius, temporalis, etc. \cite{Enoka2015}. Muscle cells contain a variable number ($N_\mathrm{m}$) of parallel myofibrils (organelles), each of which can be divided into repeated mechanical elements called sarcomeres. The typical length of a sarcomere is ca. $2 \mu$m, so there are ca. $10^5$ of these elements in series along a fiber in a typical large muscle \cite{sarcomere}. Each sarcomere contains a number of parallel thick filaments (helical bundles of myosin, red) whose constituent myosins pull on the actin polymers in the thin filaments (F-actin, blue) to generate force. Within the myofibrils, the spacing between neighbouring myosin filaments is ca. 0.046 $\mu$m at rest \cite{thick,Irving2011}. The typical cross-sectional area of a single muscle fibre substantially varies between individuals and muscle types, but is of the order of 4000 $\mu \mathrm{m}^{2}$ \cite{Lexell1991}. {Accounting for some $15\%$ of the cell volume being outside of the myofibrils \cite{Roberts2020}}, this means that a typical muscle fibre has some ca. 2,000,000 parallel filaments across, between which the macroscopic force $F$ must be divided. Rather than using this awkward number, we will express our results in terms of the {total myofibrillar CSA within a single muscle fibre}. A chemically activated muscle fibre with a CSA of 4000$\mu \mathrm{m}^{2}$ shows a force in the vicinity of 300-1000$\mu$N for untrained individuals (with a very large individual variation) \cite{Krivickas2011}, which translates to an average filament force of 150-500pN (see Supplementary A.6). {Training can increase the neural activation level \cite{Kubo2010} as well as the number of active myosin heads and the maximum voluntary contraction force per filament (\textit{e.g.} by stretch activation \cite{Hu2017}). Because of this, we would expect resistance training to lead the filament forces to tend towards the upper end of the range ($\approx 500 \mathrm{pN}$).}}
\label{fig:muscle elements}
\end{figure}

Muscles, their constituent cells, and the structure of their molecular filament mesh must respond mechanosensitively -- \textit{i.e.} in a manner which depends on the changes in the magnitude of the forces and stresses that arise during the contraction and extension of the muscle -- at many different timescales. At the fastest time scales (tens or hundreds of milliseconds), skeletal muscles can produce near maximal force for jumping or for the fight-or-flight response. Most muscles also go through cycles of shortening and lengthening with a period of the order of a second in the vast majority of sprint or endurance exercise (running, climbing, etc.) At a much longer timescale of many days, a muscle must also be able to measure changes in its overall use in order to effect adaptive muscle hypertrophy/atrophy -- ultimately helping to prevent injury on the scale of months and years.

How the muscle cell keeps track of the history of its load and stress inputs within a number of intracellular output signals (which then go on to stimulate or inhibit muscle protein synthesis), is inherently an incredibly complex biochemical question. With the help of recent theoretical insights into the folding and unfolding rates of mechanosensor proteins under force, we hope to gain insights into the first part of this puzzle for the specific case of muscle hypertrophy. 
To make progress, we use a simple model for force-induced transitions between the different conformations of the titin kinase (TK) mechanosensor. If the conformational change helps create an intracellular signal, we can model the signal's strength in terms of the duration and intensity of the mechanical inputs (external force on the TK domain in our case).

\subsection*{Force chain}

The individual sub-cellular, cellular and super-cellular components of a muscle act in concert to scale up a vast number of molecular force-generating events into a macroscopic force. The hierarchical structure of the muscle (see Fig.~\ref{fig:muscle elements}) allows the macroscopic and microscopic responses to mirror each other \citep{Uchiyama2011}.

\begin{figure}
\centering
\includegraphics[width=0.95\columnwidth]{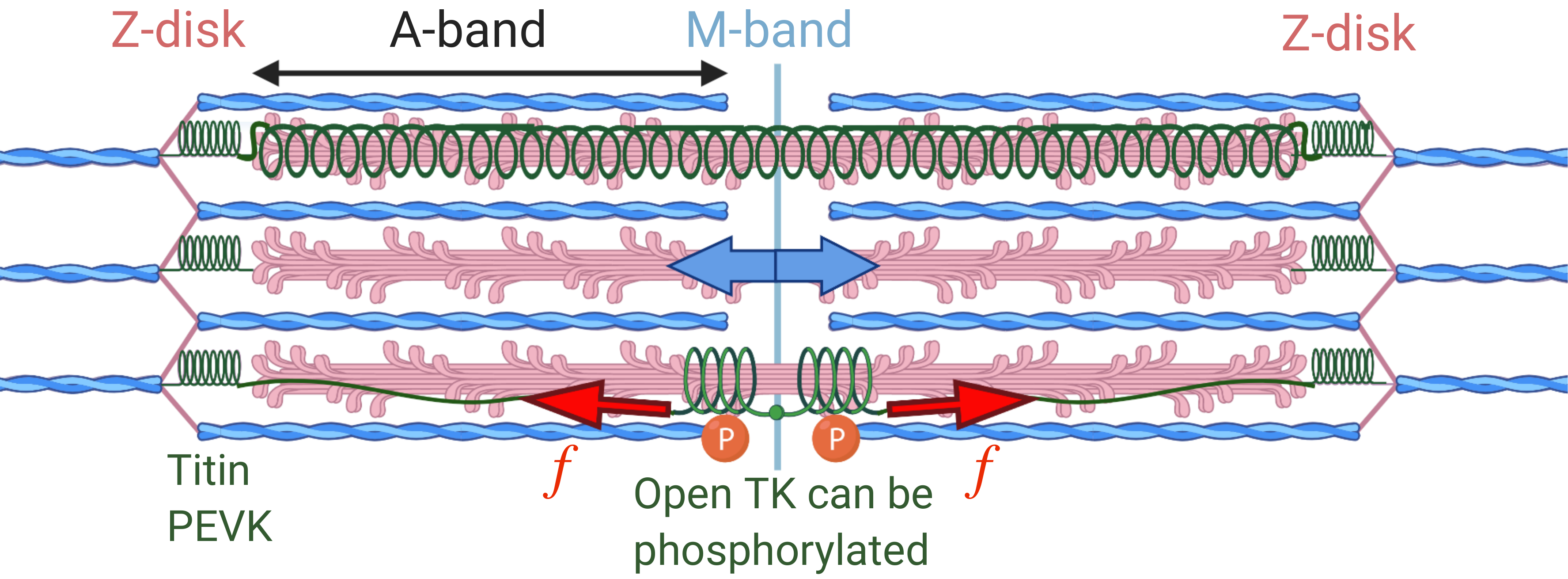}
\caption{A sketch of the mechanically active elements of sarcomere. The thick filaments are crosslinked across the M-line, with six titin molecules bonded to these filaments, on each side of the M-line. The full filament is under the measurable microscopic force identified in Fig. 1, shown by the blue arrow in the middle filament. At the molecular level, the force is borne by the individual titin and myosin filaments. { If we assume that the thick filament and titin extend by the same amount during muscle contraction, then the graphical relationship between titin force and force in the thick filament is illustrated in Figure 3 in the Supplementary Material. This figure illustrates an additional possibility: if titin wraps around the thick filament (top), then TK can lengthen substantially more than we consider in this work, for titin extending with the thick filament (bottom). The force in TK would be much higher, making TK bear more load and create a greater mechanosensitive signal.}}
\label{fig:sarcomere_sketch}
\end{figure}

The sarcomere is the elementary unit of the muscle cell and the basic building block of the sliding filament hypothesis \citep{Huxley1954,Huxley1954a}. Its regular and conserved structure, {sketched in Fig.~\ref{fig:sarcomere_sketch} for the vertebrate striated muscle}, allows for a series transmission of tension over the whole length of the muscle. In vertebrates, six titin molecules are wrapped around each thick filament \citep{AL-Khayat2013,Tonino2019} on either side of the midpoint of the sarcomere: the M-line. 

During active muscle contraction, myosin heads (motors) bind to actin and `walk' in an ATP-controlled sequence of steps \citep{Mehta2001} along the thin filaments. When a resistance is applied, the myosin motor exerts a force against it. During slow resistance training in both concentric and eccentric motions, tension is passed along the sarcomere primarily through the thin filament, myosin heads \citep{Offer1973,Knoll2012}, the thick filament, and into the cross-bridge region of the sarcomere where thick filaments are crosslinked with their associated  M-band proteins.

The load in each of the sarcomere components ultimately depends on the relative compliance of elements. The relative load on the thick filament and the M-band segments of titin when the filament is either under internal (contracting) or external (extending) load is discussed in Supplementary A.4. It is well-known that titin is under load when the sarcomere is extended \cite{Herzog2014,Hessel2017}. Recent X-ray diffraction experiments \cite{Ma2018} suggest that that the thick filament may be more compliant than originally thought; if so, M-line titin is likely substantially extended and loaded titin when the muscle actively generates force. Others disagree \cite{Reconditi2019} and attribute the change in line spacing in diffusion experiments to a mechanosensitive activation of the entire thick filament at low forces. Either way, M-band titin is under some tension during active muscle contraction. This situation is sketched in Fig.~\ref{fig:sarcomere_sketch}.

We estimate the force in each filament both macroscopically and microscopically (see the full discussion in Supplementary Part A.2 and A.3). We divide the force in the entire muscle by the number of active myofilaments (see Fig.~\ref{fig:muscle elements}) to find a large variation in force per filament in untrained individuals ($150-500$pN) \cite{Krivickas2011}. Muscle fibre neuronal and molecular activation increases with training \cite{Kubo2010,Hu2017}, so the higher forces are more likely representative of filament forces in trained individuals. The maximum filament forces extrapolated from X-ray diffraction studies \cite{Mijailovich2019} are higher at $\approx 600$pN, possibly because the actin only partially binds to myosin in normal contractions, maximum forces do not last very long, and because the muscle does not coordinate perfectly as a whole. In Supplementary Part A.4, we graphically find an approximate relation between titin and thick filament force. In particular, it suggests that the force per titin be close to $25$pN at the maximum voluntary contraction force.

In Section A.1 of the Supplementary Material, we discuss different candidates of mechanosensitive signaling in the sarcomere, and highlight the reasons why titin kinase is a particularly good candidate for this role, and why we have not considered some of these other candidates here.

\subsubsection*{TK is a mechanosensor of the `second kind'}  

Cells sense and respond to the mechanical properties of their environment using two main classes of force receptors. The first type of mechanosensor responds immediately under force \cite{Martinac2004,Takahashi2016}. Mechanosensitive ion channels are the archetypal example of such a sensor and have been proposed to play a role in tactile signaling (transforming a mechanical signal into chemical) \cite{Martinac2004,Sulivan1997}. However, the ions which they use in signaling are rapidly depleted, making it difficult for these sensors to signal in response to a sustained force.

The other type of mechanosensor, dubbed of the `2nd kind' by Cockerill et al. \cite{Cockerill2015}, can either indirectly `measure' the response coefficients, or time-integrate an external force acting on the molecule. The focal adhesion kinase (FAK) mechanosensor \cite{Bell2017,Bell2019b} is a good example: it can sense substrate stiffness by measuring the tension in the integrin-talin-actin force chain, which binds a cell to its extracellular matrix (ECM).  FAK and the TK domain both open under force, can be phosphorylated, and appear pivotal to mechanosensitive signalling, they also has many structural similarities. TK has already been suggested to act as mechanosensor \cite{Puchner2008,Tskhovrebova2008,Gautel2011}, and although recent experimental work has focussed mainly on other regions of the titin molecule, we believe that it is worth returning to the TK domain to examine it as a time-integrating mechanosensor. In the Results section below, we see that the metastability of the TK open state, when the muscle is under steady-state passive tension, can indeed allow for the TK domain to help produce increased signal levels long after the end of an exercise session. 

\subsubsection*{TK domain opens under force}

Many signaling pathways use a molecular switch to initiate a signaling cascade. One of the most common post-translational modifications of proteins involves the reversible addition of a phosphate group to some amino acids (mainly tyrosine); this addition alters the local polarity of the target protein, allowing it to change its shape and to bind a new substrate \cite{Ardito2017}. Phosphorylation can form the basis for signaling if an input changes the protein's conformation, from a native folded conformation which cannot bind to a phosphate group (often called `autoinhibited'), to an `open' conformation in which the geometry of the molecule allows phosphate groups to be donated to the phosphorylation site \cite{Bell2017}. The phosphorylated protein can then bind to a third substrate molecule, and can either directly catalytically affect or indirectly activate a signaling pathway.

Protein unfolding under force has been analysed extensively, beginning with studies of the titin Ig domain \cite{Rief1997,Oberhauser2001}. These experiments show characteristic force-extension curves, which can help deduce the transition energies between conformations for the molecules in question. We note that the Ig domains unfold under quite a high force \cite{Rief1997,Kellermayer1997,Minajeva2001} and could initially appear to be candidates for mechanosensors. However, very few phosphorylation sites have been found on the Ig domains, compared with the remainder of the molecule \cite{Hamdani2017}, suggesting that they do not contribute to force-induced signaling, but rather help control the length of the titin molecule and avoid immediate sarcomere damage under high load.

Titin kinase was initially thought to be the only catalytic domain on titin \cite{Mayans1998}. Bogomolovas et al. \cite{Bogomolovas2014} suggest that TK acts as a pseudokinase, simply scaffolding the aggregation of a protein complex when it is phosphorylated, and allowing for another protein to be allosterically phosphorylated. Computational and experimental studies of TK have shown that its force-length response also follows a characteristic stepwise unfolding pattern, but with much smaller steps than those observed for the Ig domains. In particular, AFM experiments \cite{Puchner2008} show that the presence of ATP (an energy supply) changes the conformational energy landscape of the molecule as it is stretched. This shows that the molecule possesses a long-lasting {open conformation of its TK domain}, in which it can accommodate the recruitment of signaling molecules upstream of a mechanosensitive signaling pathway, before the protein unfolds completely and potentially loses its signalling ability. Being the largest known molecule in vertebrates, titin interacts with an unsurprisingly large number of molecules \cite{Kruger2011,Attwaters2021}; {Linke et al.  \cite{Linke2008} summarized this knowledge in a protein-protein interaction network (PPIN), shown in their Figure 2, where in particular the nbr1 and MuRF pathway (localized in M-band) is shuttled into the nucleus, leading to SRF and transcription of new actin.}

\subsubsection*{Methods used in modelling}
We model a resistance training repetition as a piecewise function for force. During the loading phase (start at $t=0$), the force increases from the initial force $f(t=0)$ and asymptotically approaches the maximal force per filament $f_{\mathrm{max}}$ during the repetition, with a rate $k_f \approx 30 s^{-1}$. The full-muscle rate of force development is substantially lower, at ca. $5 s^{-1}$ \cite{Kawamori2006}, but we assume that there is a lag due to the macroscopic muscle providing some slack before macroscopic force development. It therefore seems likely that the molecular rate of sarcomere force development (which impacts the rate of titin being placed under force) is closer to the much faster rate of force increase during muscle {tetani}. During the unloading phase, the muscle force decreases with a fast rate (same rate as force development for tetani, a bit slower for twitches, but ultimately insignificant relative to the timescales of a muscle repetition). The force per titin as well as the muscle opening and closing rates $k_-$ and $k_+$ are calculated at every time step. Because the TK conformations quickly change during exercise, the next time step of the numerical integration is adaptively calculated at each time step as a fraction of the greatest fractional change in all of the molecular species in the model. Several repetitions make up a set, and several sets make up an exercise session. The exercise regime is assumed to be adaptive, such that the repetition force on TK remains constant as the total myofibrillar CSA increases.

\section{The model}

Here we explain why we believe that the kinetic processes schematically shown in Fig.~\ref{fig:reaction_diagram} are the necessary elements for any TK-based treatment of mechanosensing of the second kind and of subsequent mechanosensitive intracellular signaling.
Our model can be divided into three parts:
\begin{itemize}
\item The opening and phosphorylation of the TK domain. This stage is highly non-linear because TK opens as a mechanosensitive switch and because the mechanosensitive complex binds allosterically. The open state is metastable if the muscle is under a steady-state load.
\item The creation and degradation of signaling molecules, of new ribosomes, and of structural proteins. All of these rates can be approximated as linear, apart from a size feedback term, which arises because ribosomal diffusion is sterically hindered in large cells (see discussion below).
\item Exercise can only be so hard before the muscle depletes its short-term energy supplies. The balance between energy generation from oxidative phosphorylation and the depletion of short-term energy stores has to be considered to correctly model the dynamic response.
\end{itemize}

\subsection{Opening and phosphorylation of TK domain}

The energy barrier for the transition between the `closed' native domain conformation, and the `open' conformation which supports ATP-binding and phosphorylation is the key determinant of the kinetic transition rates between the two TK states. AFM data collected by Puchner et al. \citep{Puchner2008} is essential here; we match the relevant TK conformations to their data and explain how to extract several important model parameters in Parts A.4 and A.5 of the Supplementary Information.

\begin{figure}
\centering
\includegraphics[width=0.95\columnwidth]{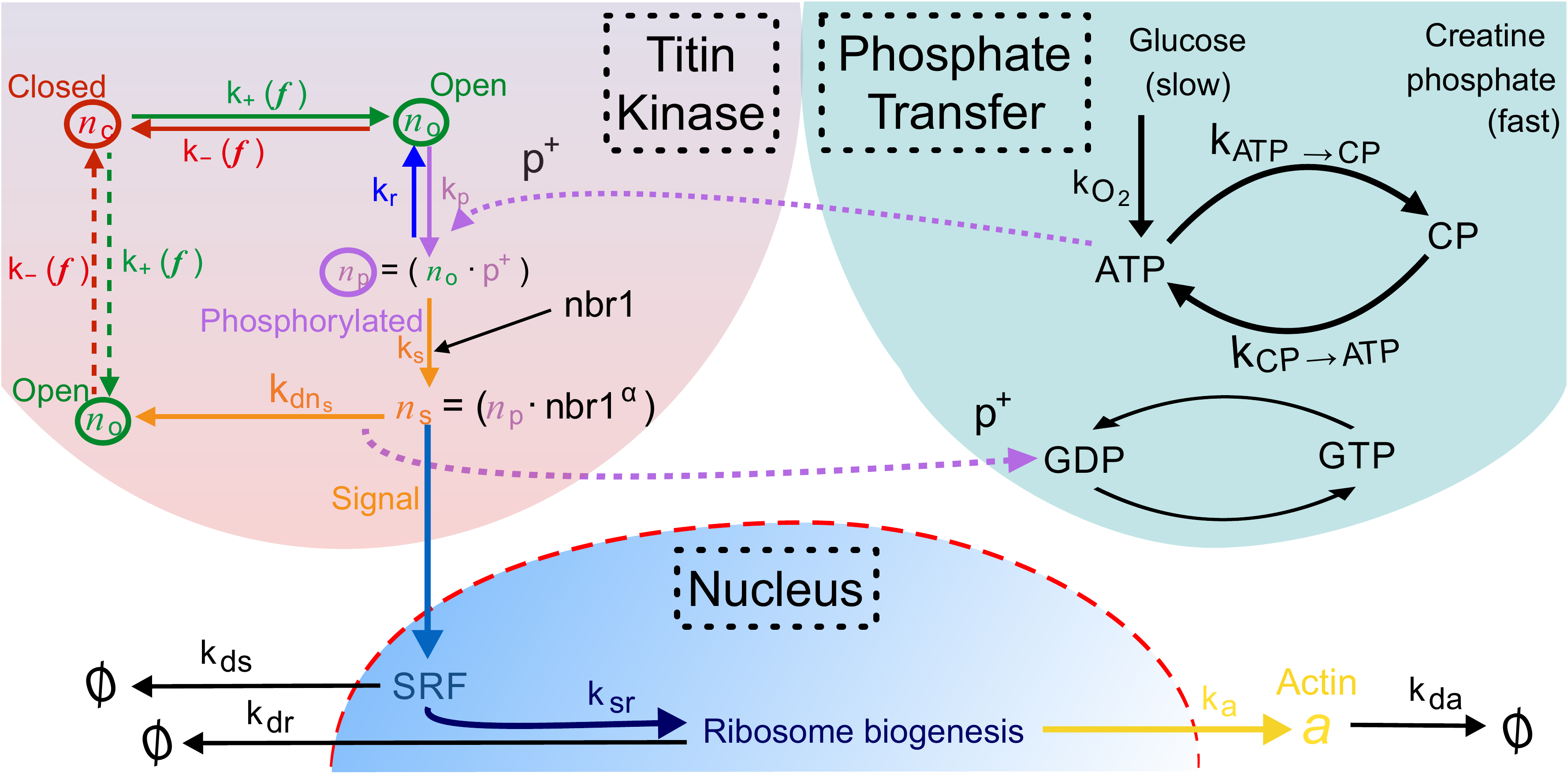}
\caption{Sketch of the kinetic processes which link titin kinase opening and phosphorylation, mechanosensing complex formation, signal activation, ribosome biogenesis and the increased synthesis of structural proteins (of these, only actin is listed for simplicity).}
\label{fig:reaction_diagram}
\end{figure}

In the absence of any signaling, the concentration of total (free+bound) ATP is constant, and the transitions from closed to open to phosphorylated TK domain conformations are simple and reversible:
\begin{itemize}
\item Closed $\leftrightarrow$ Open: TK can open under force with a force-dependent rate constant $ k_+ (f)$ and likewise close with a force-dependent rate constant $ k_- (f)$. Here we use the framework of \citep{Bell2019b} to derive these two rate constants. The concentrations of the closed and open conformations are $n_\text{c}$ and $n_\text{o}$, respectively, cf. Fig. \ref{fig:reaction_diagram}.
\item Open $\leftrightarrow$ Phosphorylated: the open state of TK can be phosphorylated with a rate constant $k_p$, the total rate of this process  depends on both the concentration of ATP and of the open TK: [ATP] and $n_\text{o}$. The phosphorylated state with the concentration $n_\text{p}$ can also spontaneously de-phosphorylate with a rate constant $k_r$, but cannot spontaneoulsy close until then.
\end{itemize}
This cyclic reaction, illustrated in the TK section of Fig. \ref{fig:reaction_diagram} is described by the kinetic equations for the evolution of $n_\text{c}$, $n_\text{o}$, and $n_\text{p}$:
\begin{align}
&\frac{d n_\text{c}}{d t} = - k_+ n_\text{c} + k_- n_\text{o} \\
&\frac{d n_\text{o}}{d t} = k_+ n_\text{c} - k_- n_\text{o} - k_p n_\text{o} \text{[ATP]} + k_r n_\text{p} \\
&\frac{d n_\text{p}}{d t} = k_p n_\text{o} \text{[ATP]}- k_r n_\text{p} \\
&n_\text{c} + n_\text{o} + n_\text{p} = n_{\text{titin}} \ \ \ \ \ \text{(constraint)} \ , 
\end{align}
where the last condition encodes the total concentration of TK units; this is equal to the concentration of titin and remains constant on the time-scale of signaling. These equations are examined in Supplementary Information, Part A.6 where they are shown to adequately reproduce the phosphorylation kinetics of TK measured by Puchner et al. \citep{Puchner2008}, providing an \textit{a posteriori} justification for their use. 

\subsection{Signal generation from phosphorylated TK}

The phosphorylated TK domain can bind the zinc finger domain protein nbr1 \cite{Lange2005}, and begin to form an aggregate; the concentration of the signaling complexes $n_s$ must be introduced with a new separate kinetic equation. The mechanosensing complex identified in the most general formulation by Lange \cite{Lange2005} is a multispecies aggregate, which we consider in more detail in Supplementary Part A.7.

SRF, the mechanosensitive signaling molecule in the Lange model, is known to undergo activation by phosphorylation \citep{Janknecht1992,Sotiropoulos1999}. There are many phosphorylation sites on nbr1, p62 and some on MuRF \cite{Hornbeck2015}, which suggests that SRF could be activated by phosphate transfer originating from TK. An activation would most likely irreversibly alter the conformation of the signaling complex, and result in the disassembly of the complex every time a new signaling molecule was activated. Assuming that the complete mechanosensing complex has a time-independent probability to disassemble, with a rate  $k_{dn_s} n_s$, we estimate the corresponding rate constant $k_{dn_s}$ from experiments \cite{Irrcher2004} that show the increase in phospho-SRF (activated signal) after exercise. They find that the level of activated SRF binding to DNA increases by a factor of 2 an hour after skeletal muscle cell contraction, and reaches half of its maximum increase after 10 minutes of exercise. This means that the degradation rate of the mechanosensing complex occurs with a half-life of ca. 10 minutes ($k_{dn_s} \approx 1/600\, \text{ s}^{-1}$).

We can now rewrite our kinetic equations to add the formation and degradation rates of the signaling complex as well as the activation of the SRF signal:
\begin{align}
&\frac{d n_\text{c}}{d t} = - k_+ n_\text{c} + k_- n_\text{o} \tag{1} \\
&\frac{d n_\text{o}}{d t} = k_+ n_\text{c} - k_- n_\text{o} - k_p n_\text{o}  \text{[ATP]}  + k_r n_\text{p} + k_{dn_s} n_s \tag{2b} \\
&\frac{d n_\text{p}}{d t} = k_p n_\text{o} \text{[ATP]}  - k_r n_\text{p} - k_s  n_\text{p} \tag{3b} \\
&\frac{d n_s}{d t} = k_s n_\text{p}- k_{dn_s} n_s \\
&n_\text{c} + n_\text{o} + (n_\text{p} + n_s) = n_{\text{titin}} \ \ (\text{constraint}) \tag{4b} \\
&\frac{d n_{\text{SRF}}}{d t} = k_{dn_s} n_s - k_{ds} n_{\text{SRF}}\ \ \ .
\end{align}
The concentration of ATP is expressed in number per titin: total phosphate is assumed to scale proportionately to the size of the myofibril and the number of titin molecules. In Supplementary Part C, we also track the kinetics of ATP depletion during intense exercise. The additional equations are mathematically more complicated, and do not help understand the full model, but are included in the numerical simulations in the Results section.

These are the core equations which describe the relatively fast activation of a signaling molecule during muscle loading. We show in the Results section below that they display a very pronounced switching behavior: in other words, small changes in tension result in large changes to the signal concentration. We also find that these equations support an increase in the concentration of signal (possibly SRF) for a substantial time of the order of a couple of days, which could help account for the immediate increase in protein synthesis post-exercise. But we shall see in the next section that a simple one-step signal cannot by itself account for the observed time-dependence of hypertrophy.

\subsection{Muscle protein synthesis after mechanosensor signaling}

The constituent molecules of most signaling pathways have a short lifetime relative to that of the structural proteins. It is also well documented that a few bouts of exercise do not have a tangible effect on muscle volume, and that muscle takes at least a few of weeks to begin to show visible hypertrophic adaptations. The debate on whether true hypertrophy is soon detected, or whether initial post-exercise changes in muscle CSA are the signs of muscle micro-damage, is a rather fraught one \cite{DeFreitas2011,Damas2016,Stock2017,Damas2018}. Three weeks of resistance training appears to be a consensus time, after which true hypertrophy is actually detected. This means that there has to be a way of `integrating' the signal over such a long period of time -- beyond the scope of the simple force-integration supported by a metastable open state of TK. Here we combine the above model of mechanosensitive signaling with a simple model of protein synthesis from a signaling molecule, and propose a mechanism by which this integration may occur.

Based on a review and discussion of the current literature in the Supplementary Information, Part B.2, we conclude that it is likely an increase in ribosome biogenesis (rather than the temporary increase in mRNA transcript number) which allows for this `time integration' of the signal. Its effect would be to suppress fluctuations in the concentration of TK conformations or signaling molecules, smoothly increasing the concentration of the structural muscle proteins over the time similar to the half-life of ribosomes. We suggest that this effect could help explain the delay of a few weeks between starting resistance exercise and the first detection of measurable muscle growth, as noted by trainers and rehabilitation specialists.

New experiments show that sarcomeric proteins are synthesised in situ at the sarcomeric Z-line and M-band \cite{Rudolph2019}. As far as we are aware, ribosomal subunits can only move by diffusion, whereas mRNA can be actively transported to the synthesis site. The inhibition of the diffusion of ribosomal subunits by the myofilament lattice \cite{Papadopoulos2000} could reduce the synthesis of new sarcomeric proteins by a sizeable amount ($5-10\%$) in adult myocytes. The fractional reduction in titin synthesis can be written in the form $-\alpha n_{\text{titin}}$, where the coefficient $\alpha$ depends on the ribosome diffusion constant, the lattice spacing and the rate of lysosomal degradation. This term has several important consequences: it provides a bound on muscle growth or shrinkage, and it affects the speed of muscle size adaptations. We examine this point in more detail in Supplementary part B.4.

We use the number of titin molecules $n_\text{titin}$ in the muscle fibre cross-section as a proxy for the muscle fibre CSA, because the hierarchical sarcomere structure is well-conserved in most muscles at rest. When necessary, one can convert from one to the other as in Fig. 1. The above equations are combined as follows (more details in Supplementary Part B):
\begin{align}
&\frac{d n_\text{c}}{d t} = - k_+(f) n_\text{c} + k_- n_\text{o} \tag{1}\\
&\frac{d n_\text{o}}{d t} = k_+ n_\text{c} - k_- n_\text{o} - k_p n_\text{o}  \text{[ATP]}  + k_r n_\text{p} + k_{dn_s} n_s \tag{2b} \\
&\frac{d n_\text{p}}{d t} = k_p n_\text{o} \text{[ATP]}  - k_r n_\text{p} - k_s  n_\text{p} \tag{3b} \\
&\frac{d n_s}{d t} = k_s  n_\text{p}- k_{dn_s} n_s \tag{5} \\
&n_\text{c} + n_\text{o} + (n_\text{p} + n_s) = n_{\text{titin}}   \tag{4b}\\
&\frac{d n_{\text{SRF}}}{d t} = k_{dn_s} n_s - k_{ds} n_{\text{SRF}} \tag{6} \\
&\frac{d n_{\text{rRNA}}}{d t} = k_{sr} n_{\text{SRF}} - k_{dr} n_{\text{rRNA}}  \\
&\frac{d n_{\text{titin}}}{d t} = k_{st} n_{\text{rRNA}} (1 - \alpha n_{\text{titin}}) - k_{dt} n_{\text{titin}} 
\end{align}
In the Supplementary Part E, we consider the possibility that the force produced by the muscle does not scale linearly with muscle size. It is unclear exactly how much active muscle force scales with muscle size. Krivickas et al. \cite{Krivickas2011} find that force increases slower at larger muscle CSA, whereas Akagi et al. \cite{Akagi2009} do not see a substantial non-linearity between force and myofibre volume. So in the main body of this paper we proceed with the simplest assumption of the linear scaling.

\begin{table}[h]
\centering
\caption{Values of rate constants, directly obtained in experiments or simulations or extrapolated from the data presented.}  \label{tab1}
\begin{tabular}{clc}
Constant & Value (s$^{-1}$) & Source \\
\hline
 $k_{p}$ & 0.07 M$^{-1}$ & \cite{Puchner2008} \\
 $k_{r}$ & 6 & \cite{Puchner2008} \\
 $k_{s}$ & $10^{-8} - 10^{-6}$ & \cite{Chen2008} \\
 $k_{dn_s} $ & 0.002 & \cite{Irrcher2004,Lange2005} \\
 $k_{ds}$ & $10^{-5}$ & \cite{Misra1991} \\
 $k_{st}$ & $10^{-5}$ & \cite{Ross1982,Ohtsuki1986,Amos1991}   \\
 $k_{dt}$ & $4\cdot 10^{-6}$ & \cite{Isaacs1989} \\
 $k_{sr}$ & 0.1 & \cite{Stoykova1983} \\
 $k_{dr}$ & ca. $9 \cdot 10^{-7}$ & \cite{Ashford1986,Ashford1986a} \\
\hline
\end{tabular}
    \end{table}

\section{Results}

The steady-state load required for the muscle to maintain homeostasis can be obtained analytically. Once we have `zeroed' our problem by checking that this value makes sense in terms of steady-state tension (muscle tone), in sections B and C, we consider the dynamics of equations (1-5) for TK only, to show that it does indeed open as a metastable mechanosensitive switch. Following that, we will proceed to study what effects different types of resistance exercise have on muscle fibre CSA, and compare them with reports from the literature.

\subsection{Steady State}

The steady-state solution to equations (1-8) is obtained in Supplementary Part D.  We find the following tension per individual TK domain:
\small
\begin{equation}
f = \frac{\Delta G_0}{u_{\text{max}}} + \frac{k_B T}{u_{\text{max}}} \ln{\Big(\frac{ \left(k_r+k_s\right)}{k_p [\text{ATP}]\big(\zeta -1 -\frac{k_s}{k_{dn_s}} -\frac{k_r+k_s}{k_p [\text{ATP}]}\big)}\Big)}
\label{eqn:steady_state_force}
\end{equation}
\normalsize
where the shorthand $\zeta$ is the ratio of synthesis to degradation coefficients:
\begin{equation}
\zeta = \frac{k_{st} (1-\alpha n_{\text{titin}}) k_s k_{sr}}{k_{dt} k_{dr} k_{ds}} \ .
\end{equation}

\begin{figure}
\centering
\includegraphics[width=0.7\columnwidth]{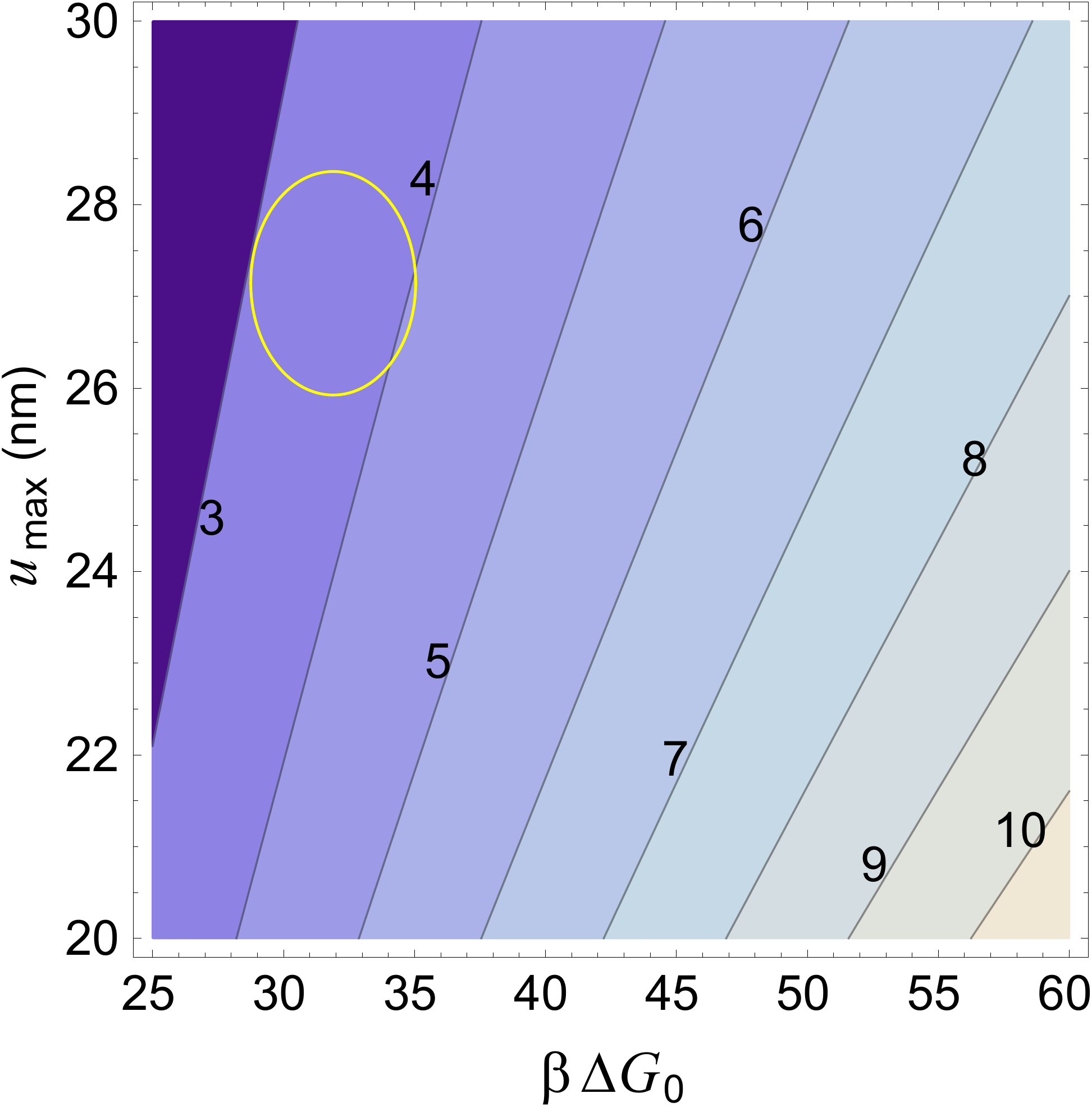}
\caption{Steady-state force (expressed in pN, labelled in contour lines) from \eqref{eqn:steady_state_force} as a function of the TK activation energy $\Delta G_0$ and the opening distance of the mechanosensor $u_{\mathrm{max}}$. $\Delta G_0$ is expressed in dimensionless units scaled by the thermal energy $\beta = 1/k_B T$, with $T=310 K$. The values of rate constants are given in Table \ref{tab1}, and the following typical concentrations were used: $p^+ = 2000$ per titin, {$\mathrm{nbr1}_{\mathrm{st}} = 0.1$ per titin, $\sigma = 0.5$, these are from the appendix and will confuse the reader}. The circle marks the `sweet spot' where the likely values of $u_{\mathrm{max}}$ and $\Delta G_0$ should be. Note that both $\Delta G_0$ and $u_{\mathrm{max}}$ are fixed physiological values, as is the maximum steady-state force. We do not have precise values for any of these (see Supplementary Part A.5 for a more detailed explanation of the uncertainty in $\Delta G_0$), so it is still instructional to plot our model's prediction of the steady-state force for different plausible values of the other two constants.}
\label{fig:resting_force}
\end{figure}

The first key result here is that the force on the TK domain, which maintains a steady state muscle fibre CSA, is determined almost exclusively by two parameters: the energy barrier $\Delta G_0$ between the closed and open conformations of the TK domain, and the unfolding distance $u_{\text{max}}$. It is clear that changing any of the coefficients in the logarithm in \eqref{eqn:steady_state_force} would only have a minor effect on the steady-state force. The typical resting muscle forces are plotted in Fig.~\ref{fig:resting_force} as a function of  $\Delta G_0$ and  $u_{\text{max}}$ (illustrated in Fig. 5 of the Supplementary Material). The typical homeostatic force experienced by a TK domain is of the order of 2-10pN.

\begin{figure}
\centering
\includegraphics[width=0.85\columnwidth]{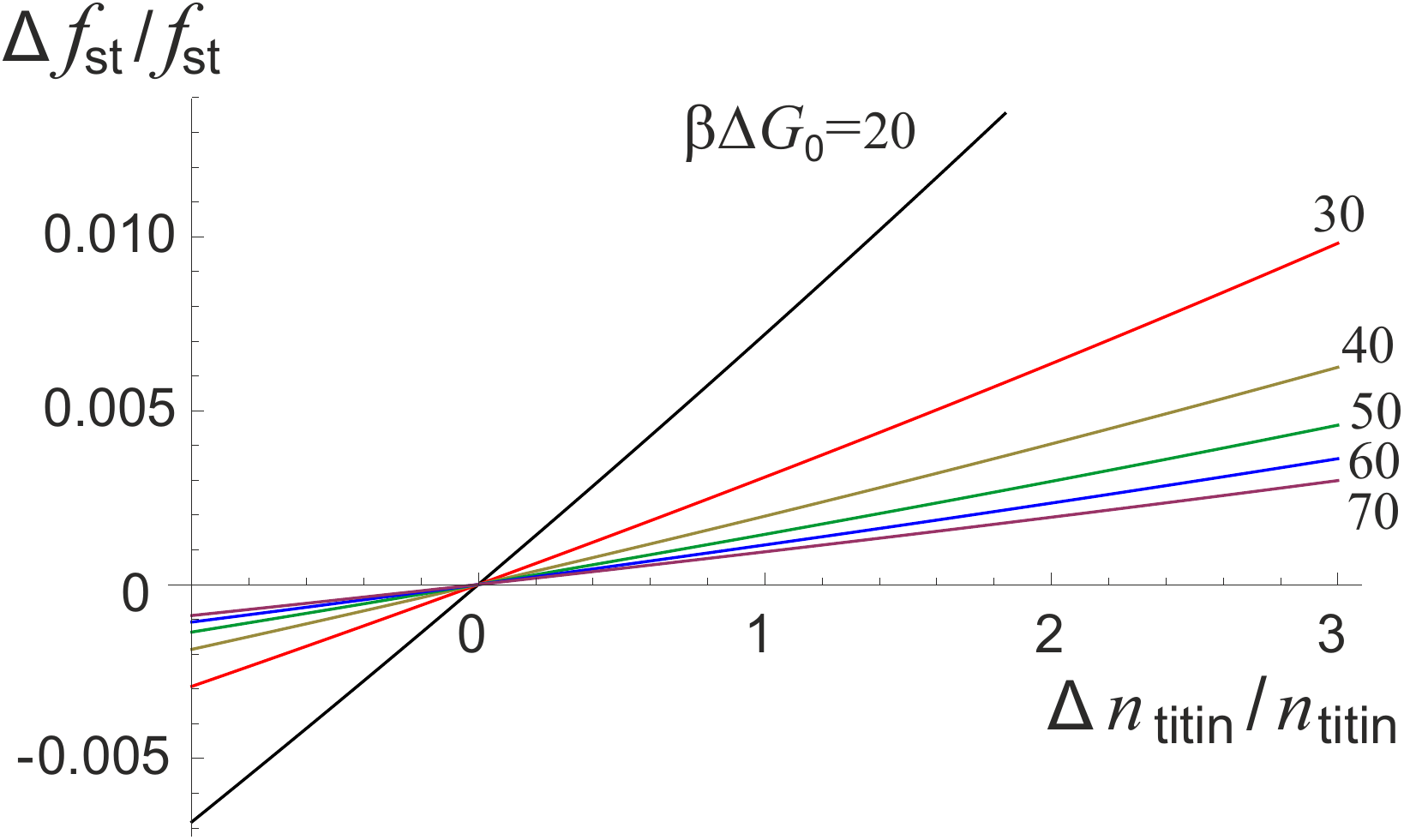}
\caption{Fractional change in steady state muscle force (vertical axis) vs fractional change in muscle size (horizontal axis), from \eqref{eqn:steady_state_force}. Note that $\Delta n_\mathrm{titin}=-n_\mathrm{titin}$ (the left limit of the axis) represents a complete degradation of the muscle. The values of opening energy $\Delta G_0$ are labelled on the plot. The values of rate constants are given in Table \ref{tab1}, and the following typical concentrations were used: $p^+ = 2000$ per titin, $\mathrm{nbr1}_{\mathrm{st}} = 0.002$ per titin, $\sigma = 0.5$. The maximum opening distance of TK was taken as $u_{\mathrm{max}} = 27 \, nm$. The values for $\Delta G_0$ and $u_{\mathrm{max}}$ were estimated from AFM data and molecular dynamics simulations conducted by Puchner et al. \cite{Puchner2008} in Supplementary Part A.5.}
\label{fig:sizetone}
\end{figure}

The other key point is that a small change in the muscle steady-state force (perhaps supported by an increase in tendon tension, which lengthens the sarcomeres) can maintain a large change in muscle size. The fractional change in the steady state muscle tone as a function of the fractional change in muscle size is plotted in Fig.~\ref{fig:sizetone}.

Combining supplementary part A.2, A.3 (maximum thick filament force) and A.4 (titin force in terms of thick filament force), we estimate the force per titin during a contraction at the maximum voluntary contraction (MVC) to be $\approx 25$pN. In the low-load regime, there is very little active muscle force (otherwise known as muscle tone), perhaps only 1-2\% of the MVC force \cite{Masi2008}  (only at most ca. 1-2pN per TK if titins were to bear most of the load), which matches well with the relative oxygen consumption in resting muscle \cite{Radak2013}. When sarcomeres operate at their optimal length, a non-negligible passive tension is developed by -- amongst other effects -- the extension of titin \cite{Moo2016}. In this regime, most of the load originates from this baseline stretch in the sarcomeres: indeed, Whitehead et al. \cite{Whitehead2001} found that passive tension at the optimal sarcomere length was of the order of 5-10\% of the MVC force. The passive tension value is much more consistent with our estimate for the resting tension per titin in the steady-state (see Fig.~\ref{fig:resting_force}).

Note that the passive force in the resting sarcomere can be substantially dialled by changing the stiffness of titin: the increased tension of the resting muscle would allow it to adjust to resistance training much more readily. The titin stiffness slowly diminishes after exercise, but the temporary increase in stiffness could also contribute to the `time-integration' of the mechanosensitive signal. This complication is beyond the scope of our model.


\subsection{Titin kinase as a metastable mechanosensitive switch}

{In Fig.~\ref{fig:nist}, we see that TK obeys switching kinetics: above a critical load, its closed conformation is no longer favoured. However, the low TK opening and closing rates $k_+$ and $k_-$ plotted in Fig.~\ref{fig:kpm} do not allow TK to quickly change between its conformations at physiological loads. If resistance exercise increases the number of open TKs, their number will remain elevated up to days after exercise; in other words, the TK open/phosphorylated/signalling complex-bound state is metastable. We use numerical simulations to explore this point further in the next section.

TK signaling increases linearly with exercise duration (barring the effects of fatigue), whereas opening rates (and signalling) increase exponentially with force in TK. {While TK force scales roughly linearly with myosin force (see Fig. 3 in the Supplementary Material), this allows mechanosensitive signaling to increase much faster than the corresponding energetic cost at high exercise force. At very high forces, however, it appears that TK force increases much more slowly than myosin force, leading to a plateau in the efficiency of mechanosensitive signalling (Fig. 4 in the Supplementary Material). Excluding mechanosensitive signalling at the steady-state force (which is efficient because thick filament force is low, but does not do much to change muscle CSA), signaling in response to resistance training is always more effective as the load increases until at least about $\approx 70\%$ of the MVC force. Our current model does not extend to how muscle fatigue induces changes in muscle stiffness \cite{Lepers2002,Chalchat2020}, which could alter TK signaling kinetics at high forces as well.}

\begin{figure}
\centering
\includegraphics[width=0.9\columnwidth]{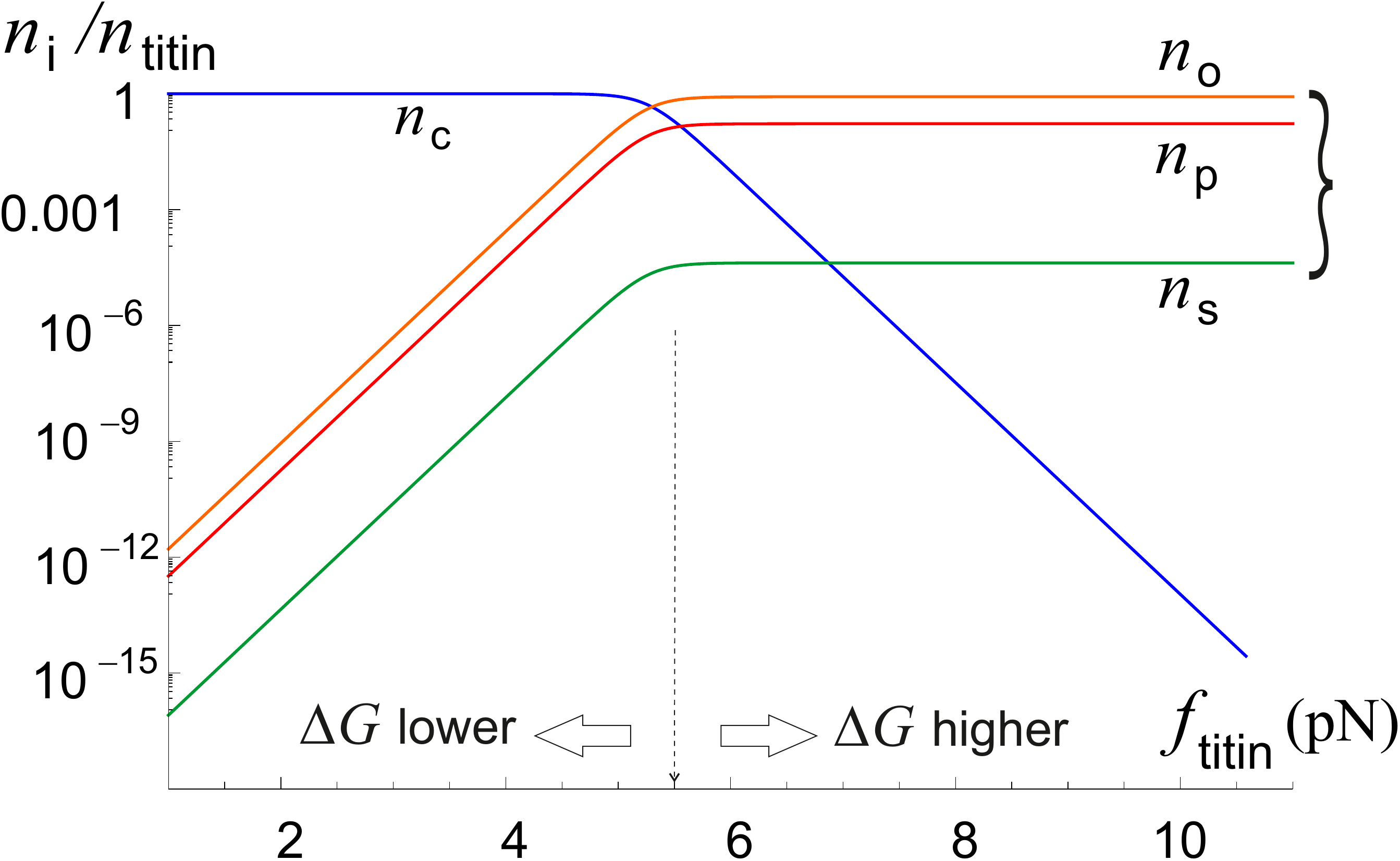}
\caption{Log-plot of the steady-state concentration of the TK conformations (blue - closed, red - open, orange - phosphorylated, green - bound to the mechanosensor complex) as a function of the steady state force per titin, from \eqref{eqn:steady_state_force}. As the steady-state force increases, the preferred conformation of TK switches from closed to a fixed ratio of open, phosphorylated and signaling complex-bound. This plot is for $\Delta G_0 = 35 k_B T$. Note that the molecule switches from being preferably closed to preferably open/phosphorylated/signaling slightly above the steady-state force of a few pN. But even though the steady-state conformation may be favoured at forces even slightly above the resting muscle tension, titin takes a long time to open enough to actually signal in large numbers, because the opening rate $k_+$ is much less than 1 s$^{-1}$ at low- and medium forces (see Fig.~\ref{fig:kpm} for an illustration of this behaviour).}
\label{fig:nist}
\end{figure}

\begin{figure}
\centering
\includegraphics[width=0.75\columnwidth]{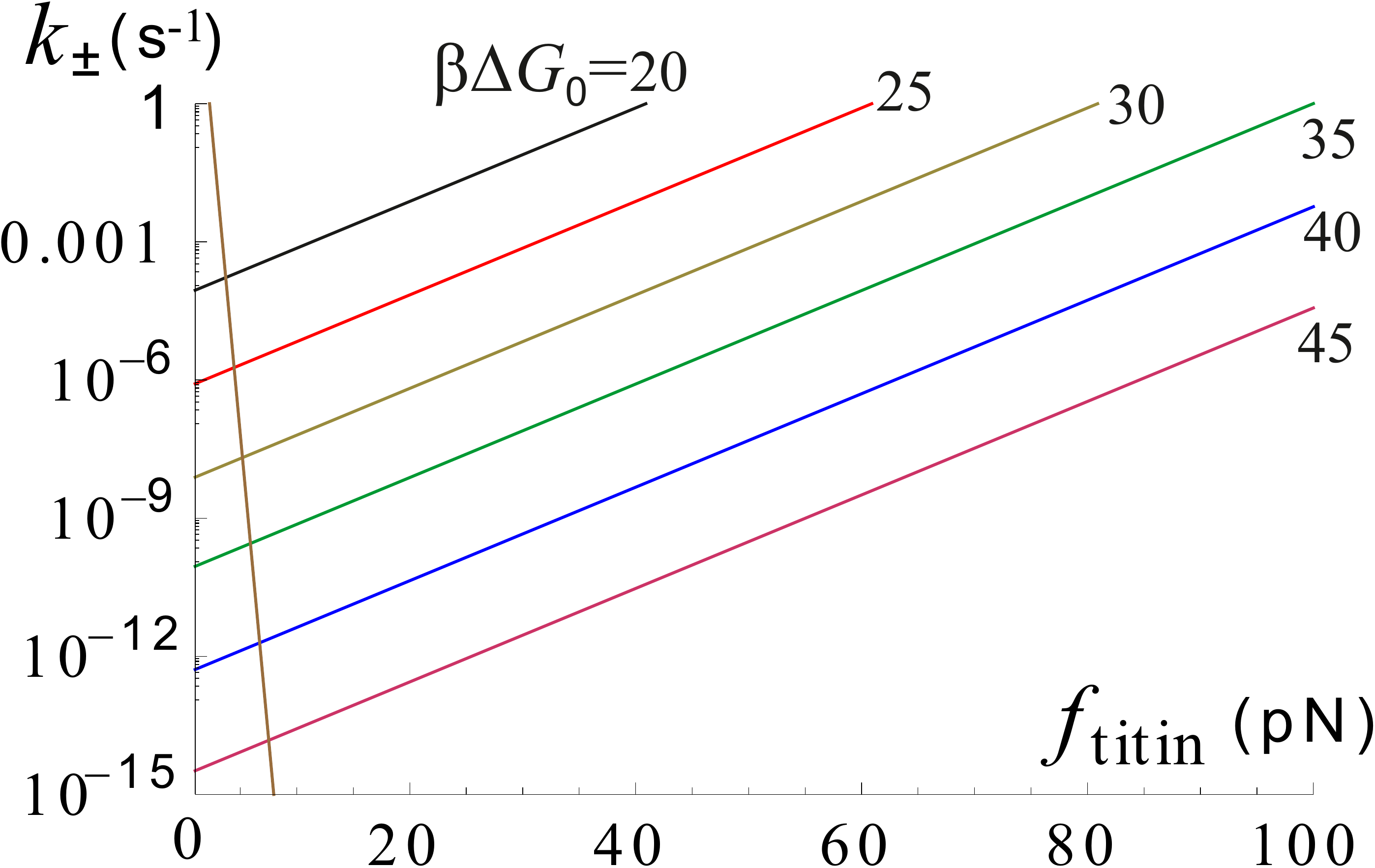}
\caption{Log-plot of the closing rate $k_-$ (brown), and opening rates $k_+$ for different values of activation barrier $\Delta G_0$, as labelled on the plot. Even when TK opening is favoured, at $k_+>k_-$, the opening rates are much less than 1 s$^{-1}$, meaning that TK opens linearly with increasing time under load, and exponentially with increasing force. We suggest that this behaviour is the basis for high intensity resistance training: doubling the force increases mechanosensitive signaling by several orders of magnitude.}
\label{fig:kpm}
\end{figure}

\subsection{Long-term mechanosensitive signaling and response}

In order to compare with experimental data in the literature, we consider a `typical' resistance exercise session consisting of 3 sets of 10 repetitions (more details in the Methods section below). This mimics a common resistance training program (see e.g. DeFreitas et al. \cite{DeFreitas2011}, who set up resistance training sessions with 8-12 repetitions to failure over 3 sets).
Choosing a specific value of repetition force is not straightforward because while most force studies consider MVC force, most hypertrophy programs compare the training load to the single-repetition maximum load for a given exercise. The muscle force during one full repetition is necessarily smaller than the instantaneous force. Determining the corresponding force per TK might be further complicated because titin is under more load when the muscle is stretched (passive force) than when it is actively contracting. Nevertheless, our choice of 20 pN per titin seems to be supported by several factors discussed here and in the Supplementary Information. 

\begin{figure}
\centering
\includegraphics[width=0.85\columnwidth]{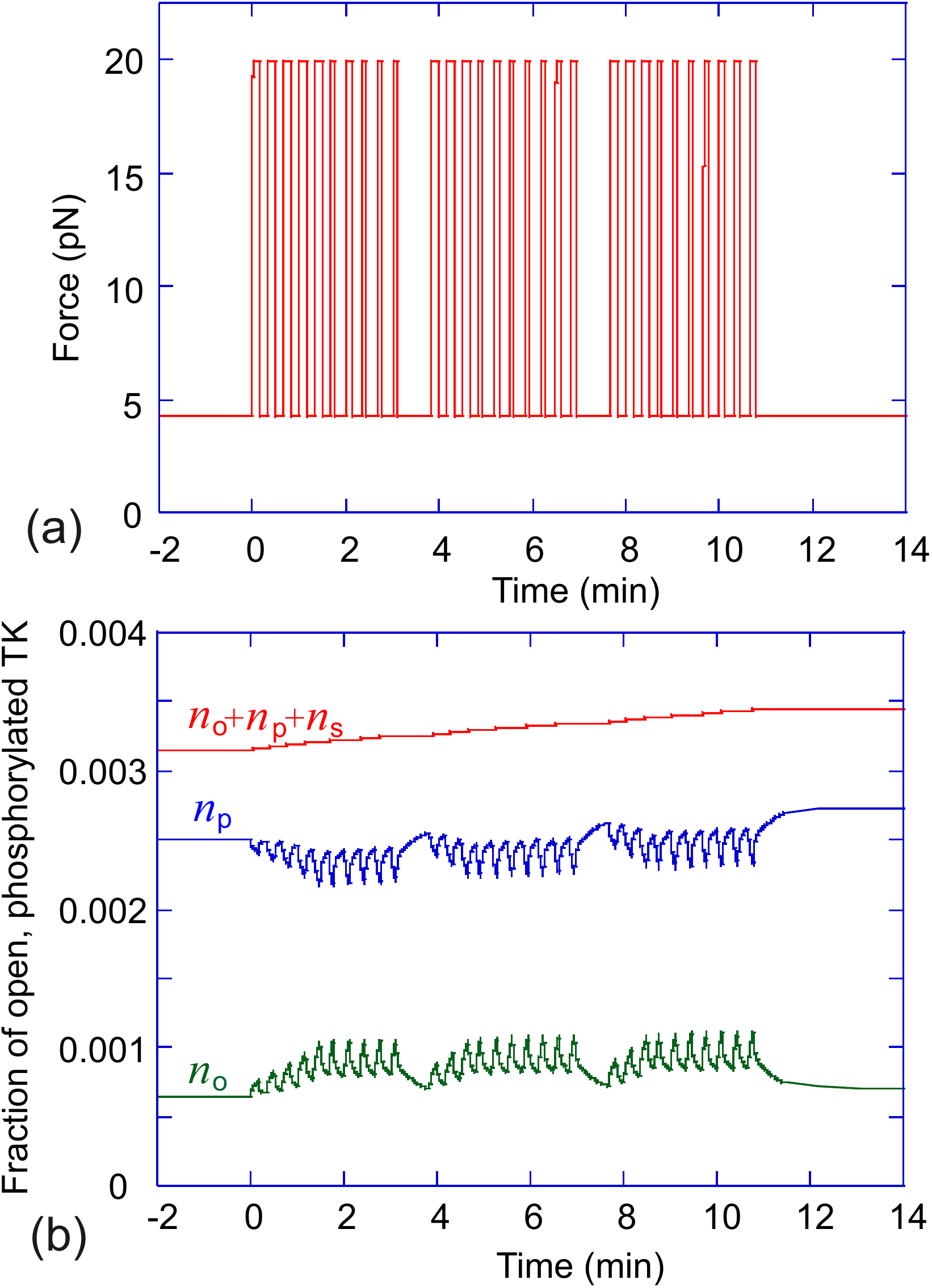}
\caption{Simulation of an exercise session involving three sets of 10 ten-second repetitions. (a) All repetitions are performed at the same force per titin, but their duration is cut short upon reaching exhaustion. As the number of titins increases, we assume that the training regime adapts by proportionately increasing the repetition force. (b) The depletion of ATP leads to a temporary drop in phosphorylated TK during exercise. However, the sum of open, phosphorylated and signaling complex-bound TK steadily increases during the exercise. Since the closing rate of TK is quite low (of the order of $10^{-5} s^{-1}$, depending on the number of attempts at crossing the energy barrier and the barrier height $\Delta G_0$), the baseline concentrations of phosphorylated and signaling TK conformations remain elevated after exercise.}
\label{fig:ex_sim}
\end{figure}

We simulate a typical exercise session as a fixed number of repetitions at a given force, grouped into a fixed number of sets, as shown in Fig.~\ref{fig:ex_sim}(a) (more details in the Methods section). During each repetition, the opening rate $k_+$ of TK becomes much greater than its closing rate, which decreases the proportion of closed TK and increases its propensity to signal. Because the muscle is under a combination of passive and active tension at rest, the closing rate of titin is small after exercise, even though it is greater than the opening rate (see Fig.~\ref{fig:kpm}). This allows TK to revert to its steady-state conformation after a time of the order of hours to days, in a manner which depends on the number of attempts at crossing the energy barrier between the closed and open conformations (see Supplementary Part A.5), as well as the height of the activation barrier $\Delta G_0$. The metastability of the open state at steady-state tension would then naturally allow the muscle to produce a mechanosensitive signal long after the end of exercise. This might account for the increase in myofibrillar protein synthesis in the two days following exercise, specifically resistance training \cite{Tang2008,Wilkinson2008}.

The important aspect of exercise, naturally reflected in our model, is the effect of fatigue. To make it more clear, we plot the same data as in Fig. \ref{fig:ex_sim}, zooming in to just one (the first) set of repetitions in Fig. \ref{fig:zoom}. Both myosin motors increase their ATP consumption under the high load, and the freshly open TK domains require ATP for phosphorylation. During the high-intensity loading, the level of ATP could drop below a critical value, after which the muscle would no longer be able to maintain the force: the only option is to drop the weight and return to the steady-state force recovery stage. We see that this effect of fatigue occurs after a few repetitons in Fig. \ref{fig:zoom}. We also find, in this simulation of model exercise, that subsequent sets of repetitions have this fatigue-driven cutoff of the later loading periods becoming less pronounced, because the overall level of ATP marginally increases during the session. 

\begin{figure}
\centering
\includegraphics[width=0.85\columnwidth]{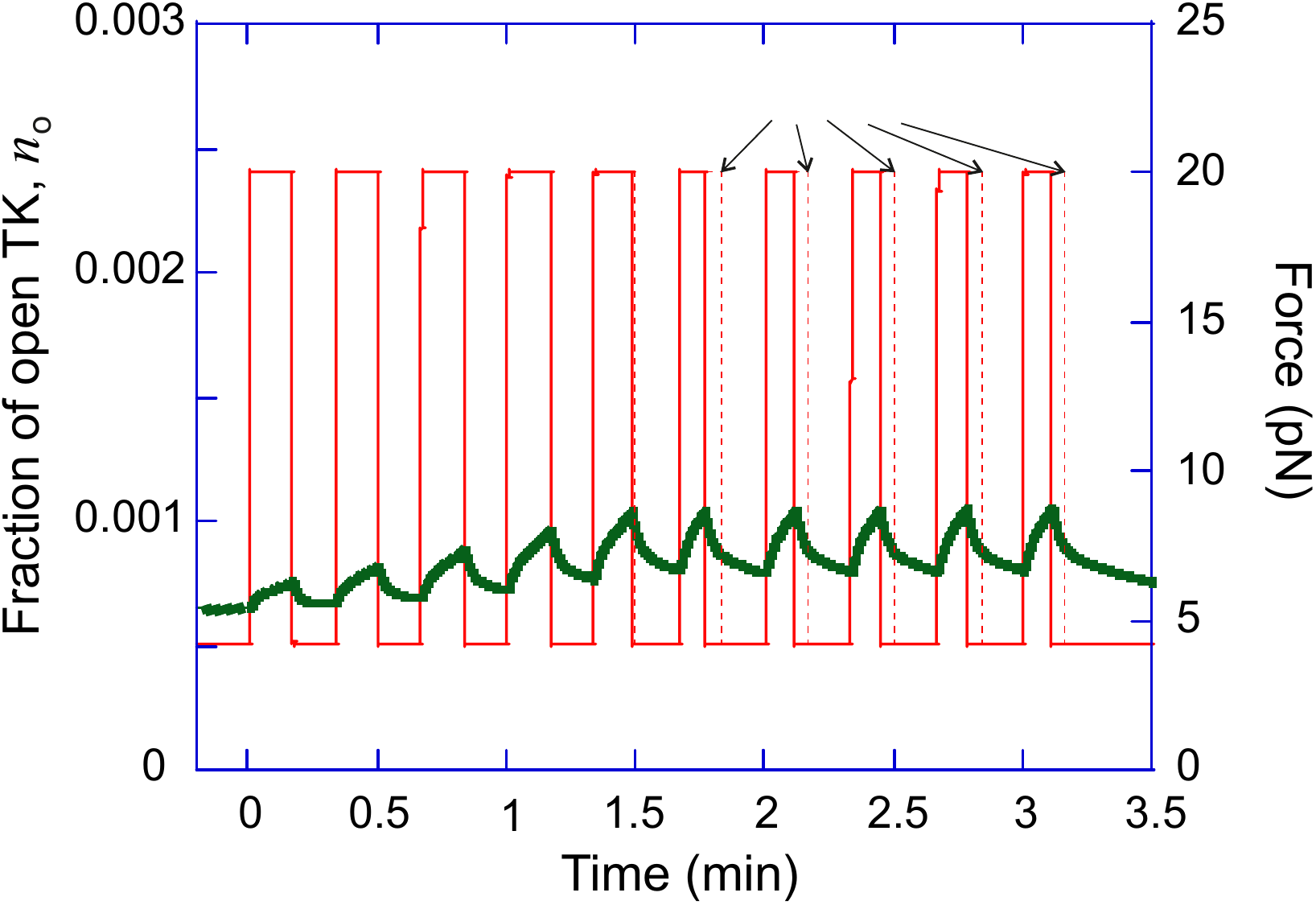}
\caption{The first set of 10 ten-second repetitions from Fig. \ref{fig:ex_sim}. Note that the repetitions become shorter as ATP runs out during the period of high load: as the ATP level falls below a critical value (which we set to a half of the homeostatic level), the muscle can no longer sustain the load and the only possibility is to drop the weight and return to the steady-state force. So the period of loading becomes shorter than the prescribed period, shon in dashed line in the plot and arrows marking the prescribed period. }
\label{fig:zoom}
\end{figure}

In Figs.~\ref{fig:hypertrophy_force} and \ref{fig:hypertrophy_freq}, and afterwards, we return to measuring the muscle `size' directly by the {total myofibrillar CSA} (by converting to that from the measure of titin molecules, which is equivalent but carries less intuitive appeal). Since the volume of a myonuclear domain is close to 16000 $\mu \mathrm{m}^3$, and remains conserved in a developed adult muscle \cite{Rosser2002}, and the density of titins is also an approximate constant (ca. 3000 per $\mu \mathrm{m}^3$), see Fig.~\ref{fig:muscle elements} -- or in an alternative equivalent estimate: the density of titins across the unit area of CSA (ca. 6000 per $\mu \mathrm{m}^2$) -- allows quantitative measure of CSA as our output.

Also note, that since in this test we are applying a constant force per titin, and the CSA increases with time, this means that the actual exercise load to the whole muscle must be increasing proportionally (in our current simplified model the relation between CSA and $n_\text{titin}$ is linear) to achieve the optimal growth.

      In Fig.~\ref{fig:hypertrophy_force} we test the long-term consequences of a regular resistance training program (the standard model exercise as in Fig.~\ref{fig:ex_sim}(a) repeated every 3 days). Several curves are presented, showing the final homeostatic saturation level, and the time to reach it, dependent on the key model parameter: the energy barrier $\Delta G_0$  for TK opening. The earlier discussion based on the data obtained by Puchner et al. \cite{Puchner2008}, and the structural analogy between TK and FAK \cite{Bell2017,Graeter2015}, suggest that $\Delta G_0$ could be around 30$k_BT$ (or ca. 75 kJ/mol). 

\begin{figure}
\centering
\includegraphics[width=0.8\columnwidth]{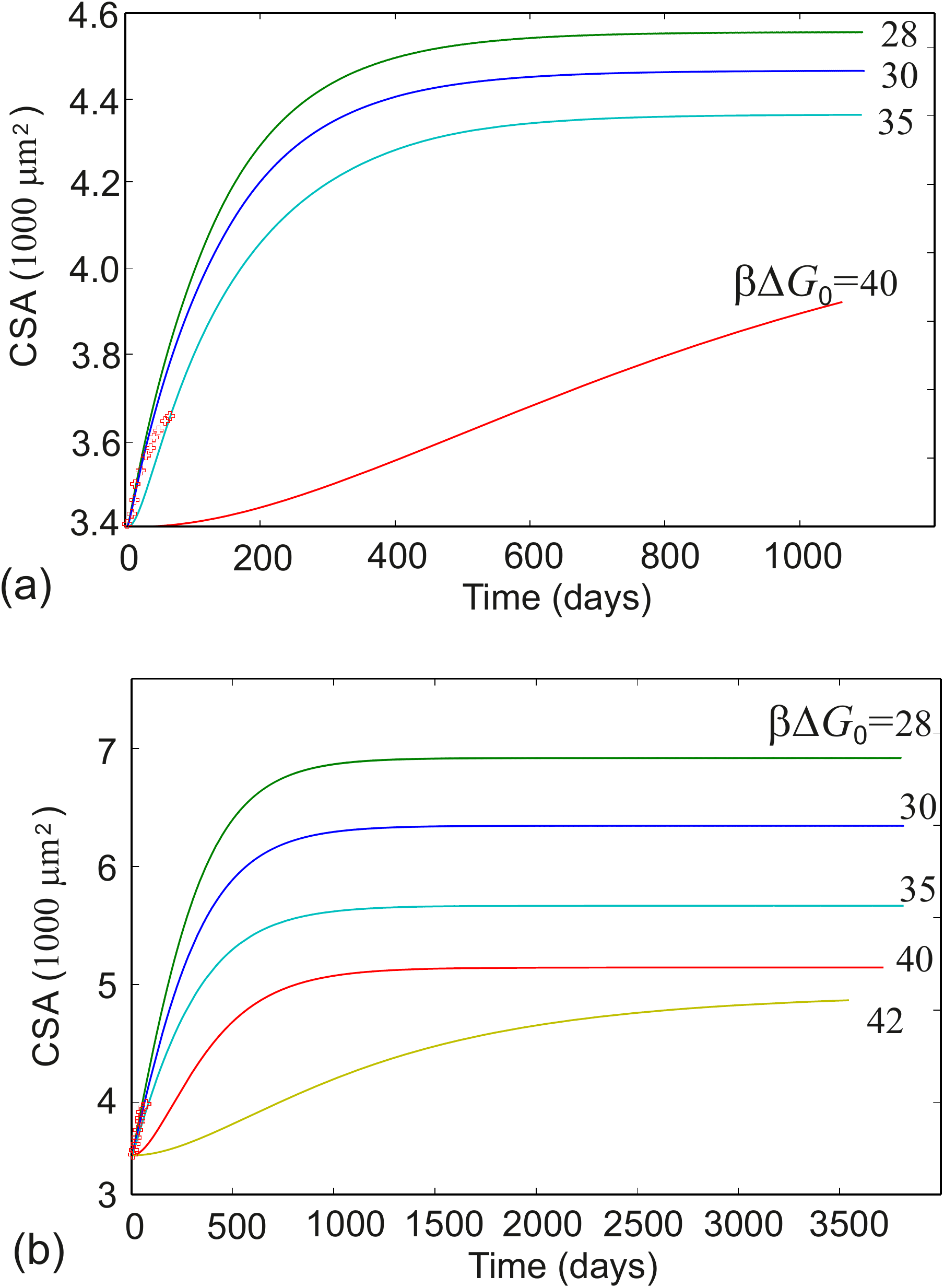}
\caption{Time course of muscle growth in response to a regular resistance training program (exercise of Fig. \ref{fig:ex_sim}, every 3 days). (a) The total muscle load $F$ is kept constant, so the force per titin $f$ effectively diminishes as the CSA increases. (b) The force per titin $f$ is maintained constant (20 pN, as discussed before), which effectively implies that the total muscle load $F$ increases in proportion with CSA (vertical axis). Several curves for different values of the energy barrier $\Delta G_0$ are labelled on the plot. As might be expected, muscle CSA changes are faster and greater in magnitude if the energy barrier $\Delta G_0$ is smaller (\textit{i.e.} TK opens faster during exercise, and signals to a greater extent). We overlay the predictions of our model with measurements of fractional changes in muscle CSA over an 8-week period, measured by De Freitas et al. \cite{DeFreitas2011} (red crosses, same values in both prlots). An initial force per titin of 20 pN matches well with real data, showing a ca. 1\% growth per week.}
\label{fig:hypertrophy_force}
\end{figure}

The comparison between plots (a) and (b) in Fig. \ref{fig:hypertrophy_force} is important. As our model relies on the value of force per titin $f$, the total load on the muscle is distributed across filaments in parallel across the CSA. So if one maintains the same exercise load, the effective force per titin diminishes in proportion to the growing CSA, the result of which is shown in plot (a). In contrast, one might modify the exercise by increasing the total load in proportion with CSA -- plot (b) shows the result of such an adaptive regime. In the non-adaptive case, the final saturation is reached in about a year and the total CSA increase is about 30\% (assuming $\Delta G_0=30k_BT$). In the adaptive exercise, the final saturation is reached much slower, but the total myofibrillar CSA increase is about 88\%: almost doubles the myofibrillar component of the muscle volume in about 2 years time. It is reassuring that the experimental measurement of De Freitas et al. \cite{DeFreitas2011} of CSA growth over a period of 8 weeks, in a similar exercise regime, quantitatively agrees with our prediction of ca. 1\% CSA increase a week in the initial period.

\begin{figure}
\centering
\includegraphics[width=0.85\columnwidth]{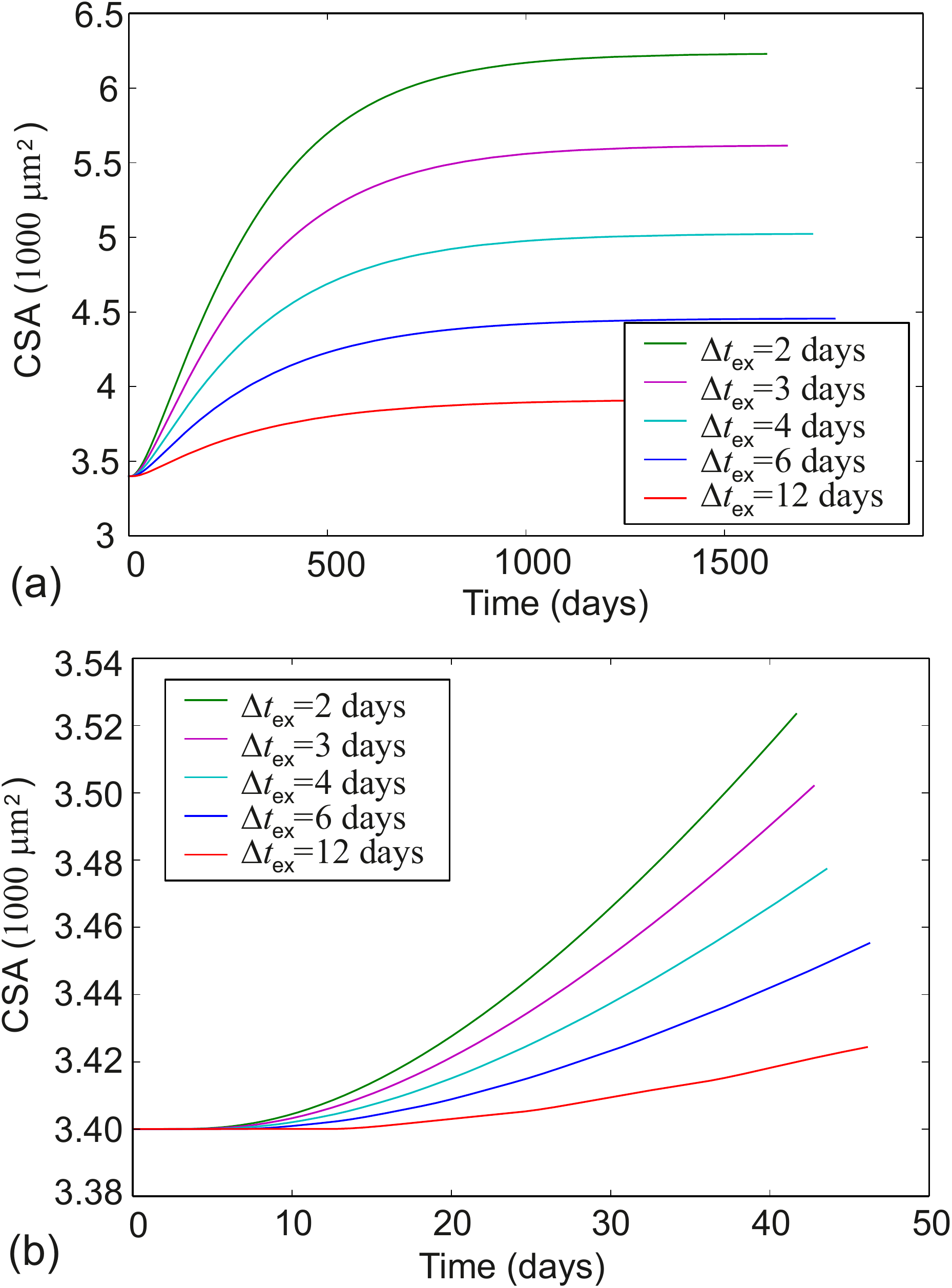}
\caption{Time course of muscle response for different exercise frequencies. Here we take $\beta \Delta G_0=35$ (see Fig.~\ref{fig:hypertrophy_force}), and a representative value of ribosome diffusion inhibition $\alpha n_{\mathrm{titin}} = 0.1$ (see discussion in Fig.~\ref{fig:detraining} for further details). (a) Full duration of the simulation: the total myofibrillar CSA asymptotically tends to a steady-state after a few years. (b) The onset of muscle hypertrophy lags the start of the exercise regime by about a week, because TK opening rates are slow, and because the signal is `integrated' by a combination of SRF and ribosomes. The initial rate of change of the total myofibrillar CSA can be compared with experiments, which show $\approx 1 \%$ CSA changes per week in response to high intensity resistance exercise \cite{Ahtiainen2003}.  This simulation shows a similar rate of CSA change, which means that a TK maximum force of $\approx 20$ pN during high-intensity resistance exercise could produce an adequate signal for muscle hypertrophy to occur. Because of the switch-like nature of TK, it is unlikely that this maximum load on TK be too different from 20 pN. This force value is consistent with a picture where the myosins bear most of the load during active muscle contraction and titin acts as a parallel stretch sensor.}
\label{fig:hypertrophy_freq}
\end{figure}

The regularity of the exercise has a strong effect: the long-term magnitude of hypertrophy predicted by the model is affected by what happens on the daily basis. Fig.~\ref{fig:hypertrophy_freq}(a) compares the long-term results when the interval between the model exercise $\Delta t_\text{ex}$ varies from frequent, to very sparse bouts (the $\Delta t_\text{ex}=$3 days case is Fig.~\ref{fig:hypertrophy_force}). We find that the extent of muscle hypertrophy is roughly linearly dependent on the exercise frequency.

We have seen that the TK mechanosensor can increase the rate of signal activation for an extended period of time after exercise. But this signal does not directly correlate with protein synthesis in the immediate aftermath of exercise. In particular, there is a known lag between the start of an exercise regime and the detection of muscle hypertrophy \cite{Stock2017,Damas2018}. This lag can be accounted for in our model if ribosomes are the main factor limiting an increase in protein synthesis, and must be made more abundant before hypertrophy can occur, see Fig.~\ref{fig:hypertrophy_freq}(b). The remainder of the results section uses the full model which includes SRF (signaling), ribosomes and titin number.

\subsection{Adaptations to resistance training exercise}

We showed in section 2.A above that constant titin kinase mechanosensing at the steady state muscle tension allows the muscle to maintain its size. In order to consider dynamic changes in muscle size, we must first assure ourselves that it reaches a new steady-state; secondly, that it predicts that muscles grow with the correct time-dependence; and finally, we must check whether the model predictions for the magnitude of change in muscle size are in the reasonable range, given that we have no free parameters (all rate constants and concentrations are independently known).

In Fig.~\ref{fig:detraining}, we see that both muscle growth during the exercise program, and muscle detraining after exercise program ends, are strongly dependent on the feedback from the slow diffusion of ribosomes across the large and sterically hindered sarcoplasm. Greater muscle fibre CSA at the start of training implies more ribosomal diffusion blocking, hence a higher hindrance term $\alpha n_{\text{titin}}$ -- resulting in a faster, lower magnitude response to the same training load. This behaviour is qualitatively observed in the literature: strength trained athletes respond to a much lesser degree to a resistance training regime, see e.g. \cite{Ahtiainen2003}. 

After stopping a resistance training programme, muscle CSA slowly decreases, eventually returning to its pre-trained homeostatic value. The time course of detraining is harder to investigate. Low values of two months \cite{Kubo2010,Coetsee2015} for skeletal muscle, to several years for recovering hypertrophic cardiac muscle \cite{Franz1998}, have been reported.
In our model, we observe reasonable time-courses which match this range for detraining for a 5-10\% degradation of ribosomes before they arrive at the sarcomere, or for very low force feedback in the range of $0.001 < \mu < 0.005$ (see Fig.~\ref{fig:detraining}).

There are some exceptions to this: career athletes maintain significantly higher muscle CSA a long time after retiring \cite{Eser2009}, and the body maintains a memory of prior resistance training events \cite{Seaborne2018} by changing its methylome. It seems likely that the body can develop and maintain a higher resting muscle tone if chronic resistance training changes the molecular architecture of the muscle. This complication is beyond the scope of our model.

\begin{figure}
\centering
\includegraphics[width=0.83\columnwidth]{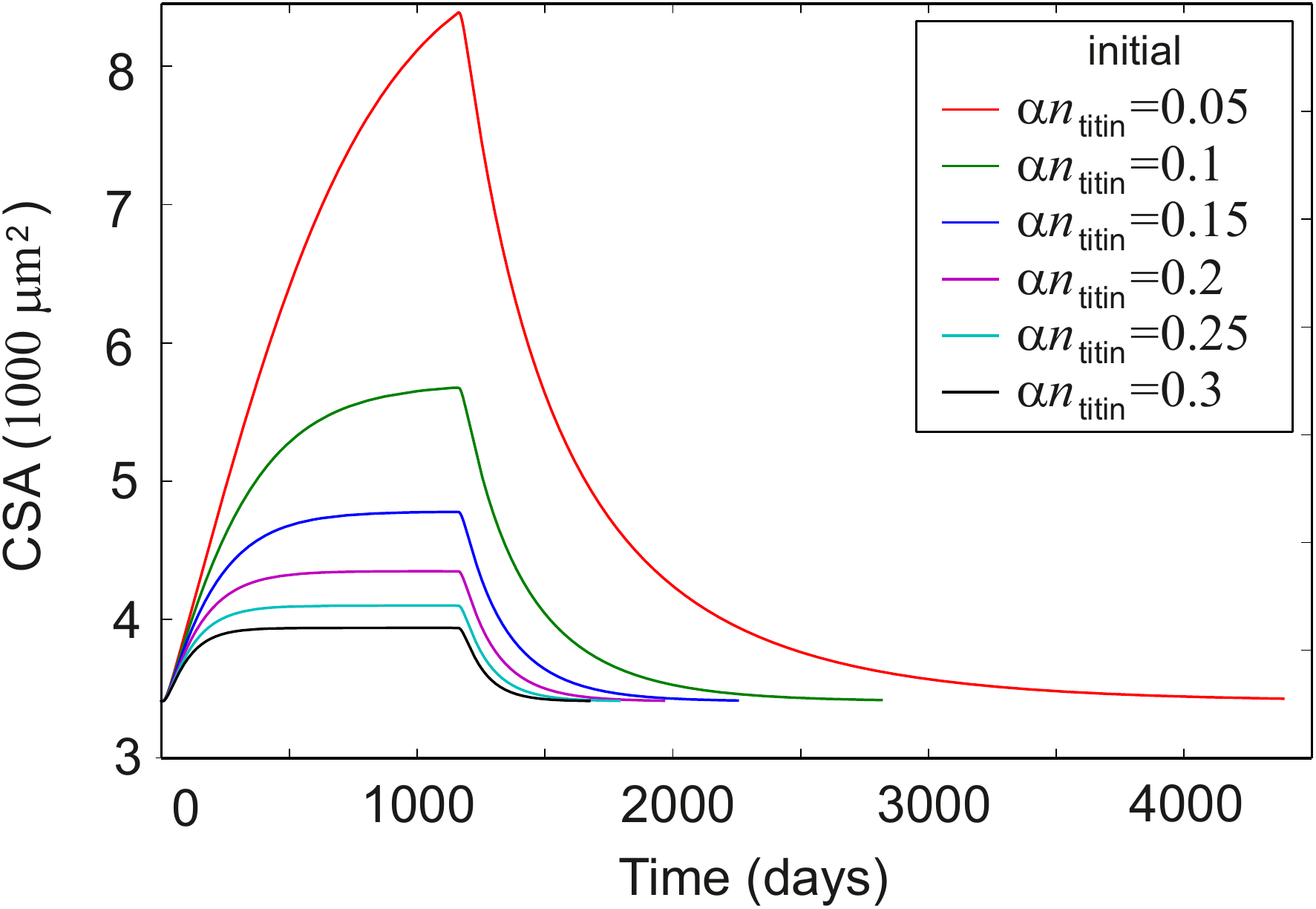}
\caption{Time course of muscle growth and loss (starting after 600 days of hypertrophy) in response to a regular resistance training program (every 3 days) with 3 sets of 10 repetitions at 20 pN per titin (our estimate of $\approx 70 \%$ 1RM), followed by detraining. The diffusive feedback depends on the degree of sarcoplasmic titin degradation, which in turn increases with myonuclear domain size and lysosomal activity. Slow detraining may combine with an initial fast loss due to atrophic conditions (see below). In this case a $5-10\%$ ribosome degradation en-route to the titin synthesis sites ($0.05 < \alpha n_{\text{titin}} < 0.1$) appears to support training and detraining at the correct rates, see \cite{DeFreitas2011,Kubo2010,Coetsee2015,Franz1998}.}
\label{fig:detraining}
\end{figure}

\subsection{Atrophy and recovery from bedrest or microgravity}

When the body is subjected to bed-rest, microgravity \cite{Parry2015}, famine \cite{Paul2012}, or as the consequence of several pathologies \cite{Bonaldo2013}, muscle size can very rapidly decrease. Any mechanism which increases degradation rates (SRF, ribosomes, titin degradation rates in our model, see Table \ref{tab1}) will necessarily cause atrophy, and our model confirms this (see Supplementary Part E for detail).

\begin{figure}
\centering
\includegraphics[width=0.9\columnwidth]{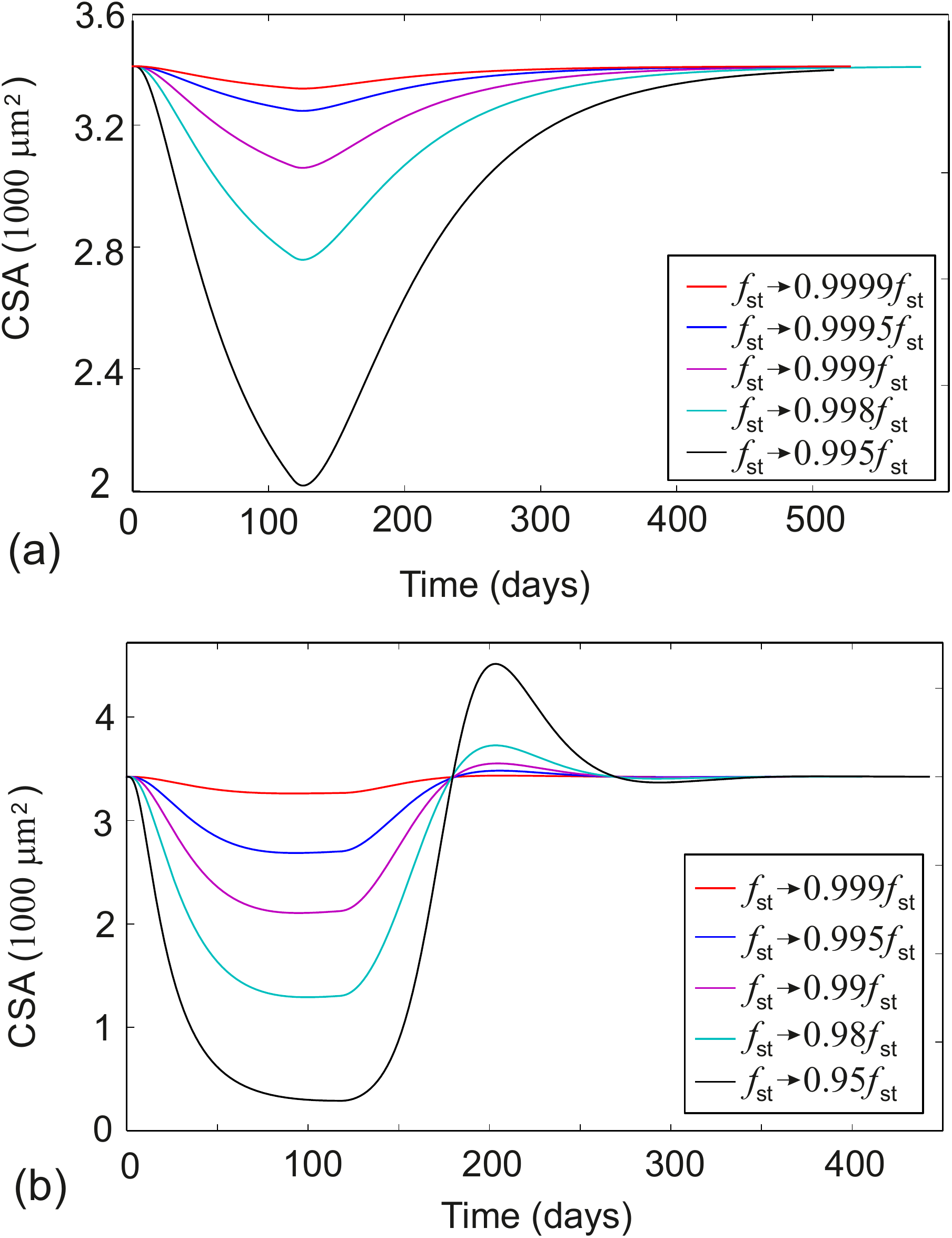}
\caption{Time course of muscle atrophy as the steady state force $f_\text{st}$ (discussed in \eqref{eqn:steady_state_force} and Fig.~\ref{fig:resting_force}) is suddenly diminished from the steady state value to a lower value. In this simulation, after 120 days, the force is brought up to its steady state value again. The recovery speed depends on exactly how the muscle force scales with muscle CSA during atrophy, the `force feedback' discussed in Supplementary Part B.4. (a) The case of negligible force feedback ($\mu = 0.005$) leads to an unphysiologically slow rates of atrophy and recovery, over several months. (b) Higher force feedback ($\mu = 0.02$) leads to a much more reasonable recovery rates, which we consider close to clinical observations.}
\label{fig:tonem}
\end{figure}

Extended periods of bed-rest and microgravity are the more interesting atrophy-inducing conditions to study in the context of mechanosensing, as it is the sudden lack of tension, which promotes muscle degradation. In other words, the steady-state force applied to the muscle (the homeostatic tone) is suddenly decreased, and the muscle metabolism responds. We find a quick decrease in muscle CSA after a series of drastic parameter changes at the start of our simulations, but it is the kinetics of muscle recovery after atrophy which appear to be more dependent on the type of feedback in the model. In practice, muscle is seen to recover relatively rapidly after very substantial atrophy, with most of the recovery occurring over a 1-2 week period \cite{Hortobagyi2000}. Fig.~\ref{fig:tonem}(a) shows our model predictions with the simplifying assumption that there was no feedback relationship between muscle force per fibre and the CSA in the case of hypertrophy. The curves show a response to a very small decrease of steady-state tone (maximum 0.5\% in black curve), and recopvery when $f_\text{st}$ returns to its value prescribed by the \eqref{eqn:steady_state_force} after 120 days. A very slow recovery of homeostatic muscle CSA is found, not in agreement with observations.  

However, once we include the feedback, when the force per filament decreases with an increasing CSA, the rate of response becomes much more realistic, see Fig.~\ref{fig:tonem}(b). Here a much greater force increase is applied (up to 5\% in the black curve), and we see both the atrophy onset and the recovery reaching the saturated steady state values within 60 days. This suggests that a reasonable force feedback scale (with the parameter $\mu \sim 0.02$ or even higher, see Supplementary Part E for detail) is a required feature of our model, if quantitative predictions are to be obtained.

An unexpected feature of plots in Fig.~\ref{fig:tonem}(b) is the muscle `overshoot' during the fast recovery after atrophy. It seems  likely that the several intrinsic processes have low rate but high sensitivity, resulting in muscle keeping a memory of its previous architecture during atrophy, much like the career-trained athletes whose muscle CSA remains higher than normal after retirement. This would translate into a corresponding increase in the muscle force at smaller muscle CSA.

\section{Discussion}

In this work, we developed a kinetic model combining the intracellular mechanosensor of the 2nd kind, the signalling chain pathway (admittedly one of several), and the ribosomal kinetics of post-transcription synthesis --  to examine how muscles sense and respond to external load patterns by producing (or degrading) their contractile proteins \cite{Attwaters2021}. The important factor of limitations to ATP supply, which affects both the MVC level due to myosin activation and the signalling due to phosphorylation, is included in the background (see Fig. \ref{fig:reaction_diagram} and Supplementary Part A). The primary marker of morphological response for us is the cross-section area (CSA) of an average muscle fibre, which is directly and linearly mapped onto the number of titin molecules per fibre. We suggest that the titin kinase (TK) domain has the right characteristics to play the role of the primary mechanosensor within the muscle cell. By looking at how TK unfolds under force, we found that it acts as a metastable switch, by opening rapidly only at high forces, but opening and closing slowly within a range of physiological forces. The muscle is known to apply a low-level tensile force, and be under a steady-state passive tension at rest, which we compare with the steady-state force predicted by our model. We find that the two forces are of the same magnitude, which suggests that long-term muscle stability is due a combination of the active muscle tone  and the passive muscle load stored in elastic sarcomere proteins -- notably titin. We find that small changes in the steady-state force allow the muscle to maintain its size after the end of a resistance training programme, and we suggest that this change in steady-state muscle tension might account for some of the `memory' that muscle develops after long-term training \cite{Eser2009,Seaborne2018}. 

Given the switch-like nature of TK, it seems likely that different individuals will have slightly different predispositions towards applying somewhat more or less muscle tone in homeostasis, and therefore can maintain muscle mass much more or much less easily. This low-level steady-state tensile force will crucially depend on the number of available myosin heads and on the steady-state ATP concentration in the cell, as well as sarcomere and tendon stiffness.

Our model shows qualitatively reasonable time courses for hypertrophy, developing during a regular exercise regime, followed by detraining as well a muscle atrophy followed by recovery. Although it is not explicitly included in the current model, long-term changes in muscle {filament} architecture (slightly increasing the muscle tone with the same CSA), as well as increases in myonuclear number after chronic hypertrophy (increasing the synthesis rates in the model), could cooperate to increase the steady-state muscle CSA. This could then provide a rationale for the observed permanent increase in muscle size after just one bout of resistance training in the past \cite{Bruusgaard2010}. The model uses no free-fitting parameters, since all its constants are independently measurable (indeed, Table \ref{tab1} gives examples of such measurements). Obviously, there would be a large individual variation between these parameter values, and so applying the quantitative model predictions to an individual is probably optimistic. However, {we are excited to develop a software} to implement the model and make specific predictions in response to any chosen `exercise regime', which could be used and adapted to practitioners.

We saw that a successive integration of the initial mechanical signal is necessary in order for muscle cells to display trophic responses at the right time scales. Indeed, a several-week lag is observed in the increase of structural proteins content after the start of an exercise regime; in our model, this arises due to a lag in ribosome number, because ribosome turnover and synthesis is relatively slow. Research into the sterically hindered diffusion of ribosomes in muscles so far appears to be very much in its infancy, despite its obvious overarching implications in muscle development. This would be an exciting avenue for future research in this area.

To further improve the model, we could include more details about the viscoelastic properties of muscle. Its effects would be twofold: first, the switching kinetics between the titin kinase conformations would change somewhat (see Supplementary Part E.2 for more details); {secondly, it would allow for a treatment of how muscle fatigue affects the compliance and therefore the mechanosensitivity of the muscle's structural proteins. We expect the plateauing of the mechanosensitive efficiency at high forces explored in Supplementary Part A.4 (in particular see Fig. S4 in the Supplementary) to be even more pronounced. In fact, because the sarcomere structural proteins likely change their mechanosensitive properties at high forces, we expect there to be an optimal force at which the exercise should be carried out.} However, this would substantially increase the mathematical complexity of the model, reducing its present intuitive clarity, and make an analytical solution for homeostasis  less tractable. This is the next stage of model development that we hope to pursue.   

In summary: how {intracellular} signaling in muscle cells organises a trophic response is a central question in exercise science and in the study of conditions, which affect muscle homeostasis (including development and ageing, as well as numerous pathologies). Cells have  been shown to use time-integrated mechanical stimuli to initiate signaling cascades, in a way which depends on the strength and duration of the signal (\textit{i.e.} mechanosensitively). This work provides a quantitative analytical rationale for a mechanosensitive mechanism for trophic signaling in muscle, and gives an additional piece of evidence that the titin kinase domain is a good candidate for hypertrophic mechanosensing. We expect advances in targeted exercise medicine to be forthcoming, specifically if the exact structure of the mechanosensing complex bound to the TK domain and its downstream signaling cascade are studied in more detail.

\begin{acknowledgments}
The authors acknowledge the significant contribution of Fionn MacPartlin, of the English Institute of Sport (www.eis2win.co.uk) in encouraging and educating us in how the muscle works in exercise and rehabilitation, and Simon Hughes, of King's College London, for critical discussions. We would also like to thank Jonathan Eddyshaw and Thomas Smith, who have contributed to the early development of this model.  This work has been funded by BBRSC DTP Cambridge (grant no. EP/M508007/1).
\end{acknowledgments}


\vspace{0.5cm}

%

\end{document}